\documentclass[11pt,titlepage]{article}
\usepackage[english]{babel}
\usepackage[dvips]{graphicx}

\begin{document}

\begin{titlepage}


\begin{center}
{\Large \bf  Charmless Final State Interaction in 
$B \to \pi \pi$ decays}\\

\vspace{1cm}

{\large \bf S. Fajfer$^{a,b,c}$, T.N. Pham$^{d}$, A. Prapotnik Brdnik$^{a,e}$\\}

\vspace{0.5cm}
{\it a) Department of Physics, University of Ljubljana, 
Jadranska 19, 1000 Ljubljana, Slovenia}\vspace{.5cm}

{\it b) J. Stefan Institute, Jamova 39, P. O. Box 300, 1001 Ljubljana, 
Slovenia}\vspace{.5cm}

{\it c) Physik Department, Technische Universit\" at M\" unchen, D-85748, Garching, Germany}\vspace{.5cm}

{\it d) Centre de Physique Teorique, Centre National de la Recherche 
Scientifique, UMR 7644, Ecole Polytechnique, 91128 Palaiseau Cedex, France}\vspace{.5cm}

{\it e) Faculty of Civil Engineering, University of Maribor, Smetanova ul. 17, 2000 Maribor, Slovenia}

\end{center}
\centerline{\large \bf ABSTRACT}

We estimate  effects of the final state interactions in $B \to \pi \pi$ decays
coming from rescattering of $\pi \pi$ via exchange of  $\rho$, 
$\sigma, f_0$ mesons. Then we include the 
$\rho \rho$ rescattering via exchange of $\pi$, $\omega$, $a_1$ mesons and 
finally we consider contributions of the  $a_1 \pi$ rescattering 
via exchange of $\rho$. The absorptive parts of amplitudes for these 
processes are determined. 
In the case of $\pi^+ \pi^-$ decay mode, 
due to model uncertainties, the calculated contribution is  
$| {\cal M}_{A}| \leq 1.7 \times 10^{-8}$ GeV. 
This produces a small relative strong phase for the tree
and color-suppressed $B \to \pi\pi$ amplitudes consistent with the result of a 
recent phenomenological analysis based on the BaBar and Belle results for 
the  $B \to \pi\pi$ branching
ratios and CP asymmetries.

\end{titlepage}
\section{INTRODUCTION}

The experimental results on B decays coming from Belle and BaBar offer many 
puzzles for  theoretical studies. 
Among them the $B\to \pi \pi$ decays are particularly interesting \cite{PDG,HFA}. 
Many theoretical frameworks such as perturbative QCD  approach of Beneke, 
Buchalla, Neubert and  Sachrajda (BBNS) 
\cite{beneke}  and the approach of \cite{sanda}, 
Soft Collinear Effective Theory (SCET) [5-9] 
and many others [10-24] 
have attempted to understand the observed decay rates.  Within QCD factorization charmless two body decays 
of B mesons have amplitudes which factorize at lowest order in $1/m_b$. 
It means that in this approach, in neglecting the next-to-leading terms in $1/m_b$ expansion, one 
ends up with the naive factorization ansatz. The naive factorization 
(e.g. \cite{WSB,ali1}) gave the rate 
of $\bar B^0  \to \pi^+ \pi^-$  too large in comparison with  the observed 
rate while the 
$\bar  B^0 \to \pi^0 \pi^0$ decay rate came out too small within this simple framework.  
Agreement with experimental data on $B \to  \pi \pi$ has been found within both BBNS and 
SCET frameworks. The improved $\bar B^0 \to \pi^0 \pi^0$ decay rate was  obtained 
recently within BBNS 
\cite{beneke} with the presence of parameter $\lambda_b$ whose precise value is unknown 
\cite{cot}. Within SCET the agreement with the experimental data is achieved \cite{SCET5} 
with the presence of  non negligible long-distance charming penguin contributions. 
It has been pointed out in  Ref. \cite{donoghue} that in B weak decays one cannot neglect
the effects of final state interactions due to the growth of forward 
scattering of the final state with the 
squared center off mass energy, as required by the optical theorem and 
cross section data. This  indicates that 
``soft scattering does not decrease for large $m_B$'' \cite{donoghue}.

Recently  the authors of \cite{cheng1}
considered two-body  decay modes by including  final state interactions (FSI).
Contributions of the $ c \bar c$  state, which in the literature 
very often called  
charming penguins were considered in  \cite{SCET5,mart}. 
The charm meson rescattering due to
charm meson exchange has been considered in Refs. 
\cite{charm-pen1,charm-pen2} and more recently in \cite{cheng1}. 
It was found the largest contribution appears 
in the $B \to K \pi$ mode \cite{charm-pen1}, but is much smaller in 
the case of $ \pi \pi$ final state 
\cite{charm-pen2}. 
The authors of \cite{cheng1} found that the absorptive part of the rescattering cannot 
explain  the observed enhancement of the 
$\pi^0 \pi^0$ branching ratio and cannot produce  
a small branching ratio of the $\pi^+ \pi^-$ rate.

Motivated by this study \cite{cheng1}  we reexamine 
final state interactions 
in $ \bar B^0 \to \pi^+ \pi^-$ and $ \bar B^0 \to \pi^0 \pi^0$ modes which 
result from the light mesons rescattering. 
  We use  mainly the same framework
as described in \cite{cheng1}, but we point out that there are more intermediate states which 
contribute to both amplitudes and give important  contributions. 
As in \cite{cheng1} we take into account 
only dominant contributions proportional to the effective Wilson coefficient $a_1$. In this 
approach for the charmless final state interactions only the contributions of 
$\pi \pi$ and $\rho \rho$ intermediate states were used in \cite{cheng1}. Since in B decays, 
 resonant FSI is expected to be suppressed due to the
absence of resonances at energies close to the mass of the 
B meson, we  consider only $t$- channel
FSI. However, in the case of $\pi \pi \to \pi \pi$ rescattering  we include 
 possibility that in addition to the $\rho$ meson exchange there are  
 contributions coming from $\sigma$ and $f_0$ exchange. 
In the case of $\rho \rho \to \pi \pi$ 
rescattering we find that there is a contribution of the  $\omega$ meson for 
the $\pi^+ \pi^-$ final state  
as well as contributions of the $a_1(1260)$ axial meson. 
We determine contributions coming from $a_1(1260) \pi$ intermediate 
states, inspired by the recent BaBar 
measurement of the very large rate for 
$\bar B^0 \to a_1^- \pi^+$ state with the branching ratio 
${\rm BR}(B^0 \to a_1^+ (1260) \pi^-)$ 
$  = (40.2 \pm 3.9 \pm 3.9) \times 10^{-6}$ \cite{BaBar}. In our approach the $a_1(1260)^- \pi^+$
rescatter via $\rho^0$ exchange into the $\pi^+ \pi^-$ final state. 
Although the $\bar B^0 \to a_1^+ \pi^-$ decay rate has not been observed yet, 
we estimate this contribution assuming the 
naive factorization for the amplitude. 
The paper is organized as follows: in Sec. 2 we give basic formulas for 
the two-body B 
amplitudes and the Lagrangian describing the strong interactions of the 
light mesons used in our calculations, in Sec. 3 
we present results of our calculations for the absorptive part of the 
amplitude, in Sec. 4 we discuss our results and  we summarize them in Sec. 5.  

\section{THE FRAMEWORK}

In the studies of  
the $B \to \pi \pi$ and $
B \to K \pi$ branching ratios and CP 
asymmetries it was found that  amplitudes arise from 
tree,  color-suppressed, penguin and the electroweak penguin 
diagrams  (see e.g. \cite{gronau,cheng1}). 
In our approach we consider  only leading contributions in charmless FSI 
 and therefore we only use    the 
effective weak Lagrangian for the process $b  \to \bar u d  u$ at the tree level 
in the following form:
\begin{equation}
{\cal L}_w  = -\frac{G}{\sqrt{2}} V_{ub} V_{ud}^* a_1 ( \bar u b)_{V-A}
(\bar d u)_{V-A}.
\label{e1}
\end{equation}
Here $a_1$ is the  Wilson coefficient and we use the same value as given 
in \cite{cheng1} ($a_1 (\mu) = 0.991 + i 0.0369$; the scale $\mu =2.1$ GeV), 
which includes short-distance nonfactorizable 
corrections such as vertex corrections and the hard spectator interactions 
 determined within QCD factorization approach \cite{beneke}. 
In our further study we use 
naive factorization approximation \cite{WSB}, in which the B meson decay 
amplitude can be written 
as a product of two weak current matrix elements.  
The standard decomposition of the weak current matrix elements is: 
$$
\langle V(k,\varepsilon,m_V)|\bar q\Gamma^\mu
q|P(p,M)\rangle=\epsilon^{\mu\nu\alpha\beta}\varepsilon_\nu
p_\alpha k_\beta \frac{2V(q^2)}{M+m_V}+2im_V\frac{\varepsilon \cdot
q}{q^2}q^\mu A_0(q^2)
$$
\begin{equation}
+i(M+m_V)\Big[\varepsilon^\mu-\frac{\varepsilon
\cdot
q}{q^2}q^\mu\Big]A_1(q^2)-i\frac{\varepsilon\cdot q}{M+m_V}\Big[P^\mu-\frac{M^2-m_
V^2}{q^2}q^\mu\Big]
A_2(q^2)\;.
\label{PvV}
\end{equation}
Similarly, heavy pseudoscalar to light pseudoscalar transition is described 
by the matrix element:
\begin{equation}
\langle P(k,m_P)|\bar q\Gamma^\mu q|P(p,M)\rangle=
\left[P^\mu-\frac{ (M^2-m_P^2)}{q^2}q^\mu\right]F_+
(q^2)+\frac{(M^2-m_P^2)}{q^2} q^\mu F_0(q^2)\;,
\end{equation}
 while for the heavy pseudoscalar to light axial vector transition, we use the expression
given in \cite{cheng-f+}: 
$$
\langle A(k,\varepsilon,m_A)|\bar q\Gamma^\mu
q|P(p,M)\rangle= i \big [ (M +m_A) \varepsilon^\mu  V_1(q^2)  - \frac{\varepsilon \cdot
q}{M +m_A} P^\mu V_2(q^2) -$$
\begin{equation}
2 m \frac{\varepsilon \cdot q}{q^2} q^\mu (V_3(q^2) - V_0(q^2)) \big ]
- \epsilon^{\mu\nu\alpha\beta}\varepsilon_\nu
p_\alpha k_\beta \frac{2A(q^2)}{M+m_A}\,,
\label{PA}
\end{equation} 
with $V_3 (q^2) = (M +M_a)/(2m_A) V_1(q^2) - (M-m_A)/(2m_A) V_2(q^2)$. 
In above equations $q^\mu=p^\mu-k^\mu$ and $P^\mu=p^\mu+k^\mu$. 
The light meson creation (annihilation) is described by the matrix elements:
\begin{eqnarray}
&&\langle P(p)|\bar q\gamma^\mu(1-\gamma_5)q|0\rangle=if_P p^\mu\;,
\qquad
\langle V(p,\varepsilon) |\bar q\gamma^\mu(1-\gamma_5)q|0\rangle=
f_V m_V\varepsilon^\mu\;,
\nonumber \\
&&\langle A(p,\varepsilon)|\bar q\gamma^\mu (1-\gamma_5) q|0\rangle  
=f_A m_A \varepsilon^\mu\;.
\label{fconst}
\end{eqnarray}
In our numerical calculations we use the following values of relevant
parameters as given in \cite{cheng1}: 
$f_{\pi} = 0.132$ GeV, $f_{\rho} = 0.21$ GeV,  $f_{a_1} =0.205$ GeV, $F_0^{B\pi}(0) \simeq 
F_0^{B\pi} (m_{\pi}^2) = 0.25 $ $\simeq F_+^{B\pi} (m_{a_1}^2)$, $A_1^{B\rho}(0) \simeq 
A_1^{B\rho} (m_{\rho}^2) = 0.27$, $A_2^{B\rho}(0) \simeq 
A_2^{B\rho} (m_{\rho}^2) = 0.26$. We use:  $V_0^{B a1} (0) \simeq 
V_0^{Ba1} (m_{\pi}^2)$ $=0.13$ \cite{cheng-f+}.

Using above expressions 
the leading contribution to the   amplitude for $\bar B^0 \to \pi^- \pi^+$ was found to be 
(e.g. \cite{ali1})
\begin{eqnarray}
&&{\cal A}(\bar B^0 \to \pi^+\pi^-)=i{\cal A}_\pi =-i \frac{G}{\sqrt{2}} V_{ub} V_{ud}^* a_1 [F_0^{B\pi}
(m_\pi^2)(m_B^2 - m_\pi^2)] f_\pi\,. 
\label{e2p}
\end{eqnarray}
In \cite{cheng1} the value $a_1 = 0.9921 + i 0.036$ 
led to the amplitude  $ {\cal A}_\pi (\bar B^0 \to \pi^+\pi^-)_{\rm SD} =
3.2\times 10^{-8} + i 1.2 \times 10^{-9}$ GeV 
(we took the $V_{ub} =0.00439$ \cite{uli}). 
Without color-suppressed and penguin contributions this gives the branching ratio  
$BR(\bar B^0 \to \pi^+\pi^-)_{\rm SD} = 9 \times 10^{-6}$, too large in comparison with the average experimental 
value $(4.6 \pm 0.4) \times 10^{-6}$ as given in \cite{cheng1}. The 
inclusion of color-suppressed  and penguin amplitudes decreases the rate \cite{ali1,cheng1}, 
but it is still too large 
in comparison with experimental result.

The amplitude for $\bar B^0 \to \rho^+  \rho^-$ is 
$${\cal A}(\bar B^0 (p_B) \to \rho^+(q_1, \epsilon_1)  \rho^-(q_2,
\epsilon_2)) =  i {\cal A}_\rho
\left(\epsilon^{\mu \nu \alpha \beta} \epsilon_{1 \mu}\epsilon_{2 \nu}q_{1\alpha}
  q_{2\beta} \frac{-2i V(m_\rho^2)}{M_B +m_\rho} \right. 
$$
\begin{equation}
\left. + A_1(m_\rho^2) (M_B +m_\rho) \epsilon_{1} \cdot \epsilon_{2} -  
2 A_2(m_\rho^2)\frac{\epsilon_1 \cdot p_B  p_B \cdot \epsilon_2}{M_B +m_\rho} \right)\,,
\label{e7a}
\end{equation}
with $A_\rho =-\frac{G}{\sqrt{2}} V_{ub} V_{ud}^* a_1 f_{\rho} m_{\rho}$. 
The amplitudes for $\bar B^0 \to a_1^- \pi^+$ and $\bar B^0 \to a_1^+ \pi^-$  are:
\begin{eqnarray}
&&{\cal A}(\bar B^0(p) \to a_1^-(q_2, \epsilon) \pi^+(q_1)) = 
i{\cal A}_{a_1,1} (p +q_1) \cdot\epsilon,
\nonumber \\ &&
{\cal A}(\bar B^0(p) \to a_1^+(q_1, \epsilon) \pi^-(q_2)) = i{\cal A}_{a_1,2}
 (p +q_1) \cdot\epsilon, 
\label{e7}
\end{eqnarray}
with ${\cal A}_{a_1,1} = -\frac{G}{\sqrt{2}} V_{ub} V_{ud}^* a_1 f_{a_1} m_{a_1} 
F^{B\pi}_+(m_{a1}^2)$ and
${\cal A}_{a_1,2} = -\frac{G}{\sqrt{2}} V_{ub} V_{ud}^* a_1f_{\pi} 
2 m_{a_1}V_0^{Ba1}(m_\pi^2)$.

The light mesons' strong interactions are described by 
\begin{eqnarray}
{\cal L}_{strong}& =& i \frac{g_{\rho \pi \pi}}{\sqrt2} Tr( \rho^\mu
[\Pi, \partial_\mu \Pi]) - 4 \frac{C_{VVP}}{f} \epsilon^{\mu \nu \alpha
  \beta}
Tr (\partial_{\mu} \rho_{\nu}  \partial_{\alpha}
\rho_{\beta} \Pi)\nonumber\\
&+& G_{AVP}Tr (A_{\mu}
[ \rho^\mu, \Pi]) + iG_s {\sqrt 2}Tr (\Pi \Pi S)+ iG_{s'}{\sqrt 2} Tr (\Pi \Pi S')\,.
\label{e4}
\end{eqnarray}
In these equations $\Pi$ is the $ 3 \times 3$ matrix containing
pseudoscalar
mesons, $\rho$ is the $ 3 \times 3$ matrix describing light vector mesons,
and $S$, $S'$ are matrices describing scalar mesons. 
In our numerical calculations we use  $g_{\rho \pi \pi}= 5.9$ and $C_{VVP}= 0.33$ (see \cite{bando} -
\cite{bramon}).
The coupling $|G_{AVP}| = 3.12$ GeV is obtained from the 
experimental results for $a_1^0 \to \rho^- \pi^+$ decay width 
$\Gamma_A = 0.2$ GeV. 
Finally, the couplings $G_s$ and $G^\prime_s$ are obtained by using 
PDG data \cite{PDG} on $\sigma$ (or $f_0 (600)$) and $f_0(980)$ meson: $m_{\sigma} \approx (0.4 - 1.2)$ GeV,
$\Gamma_\sigma \approx (0.6 -1)$ GeV,  $m_f= 0.98$ GeV
and $\Gamma_f \approx (0.04 - 0.1)$ GeV. In the numerical calculation we take
the average values 
$m_{\sigma}= 0.8$, $\Gamma (\sigma \to \pi\pi) = 0.8$ GeV, 
$m_f= 0.98$ GeV and   $\Gamma (f_0(980) \to \pi \pi) = 0.07$ GeV  
and we determine $G_s = 4.24$ GeV, and $G_s^{\prime} = 1.37$ GeV. 

Using naive factorization we obtain for the  
branching ratio ${\rm BR}(\bar B^0 \to a_1^- \pi^+) =1.8 \times 10^{-5}$ about 
two times smaller than the experimental result  given in \cite{BaBar}. 
Using above mentioned data we predict 
that ${\rm BR}(\bar B^0 \to a_1^+ \pi^-) = 8.2 \times 10^{-6}$.

\section{THE ABSORPTIVE PARTS OF THE AMPLITUDES}

In our calculation of the absorptive parts of amplitudes we include 
the contributions coming from the graphs presented in Fig. 1.
\begin{figure}
\includegraphics[width=17cm]{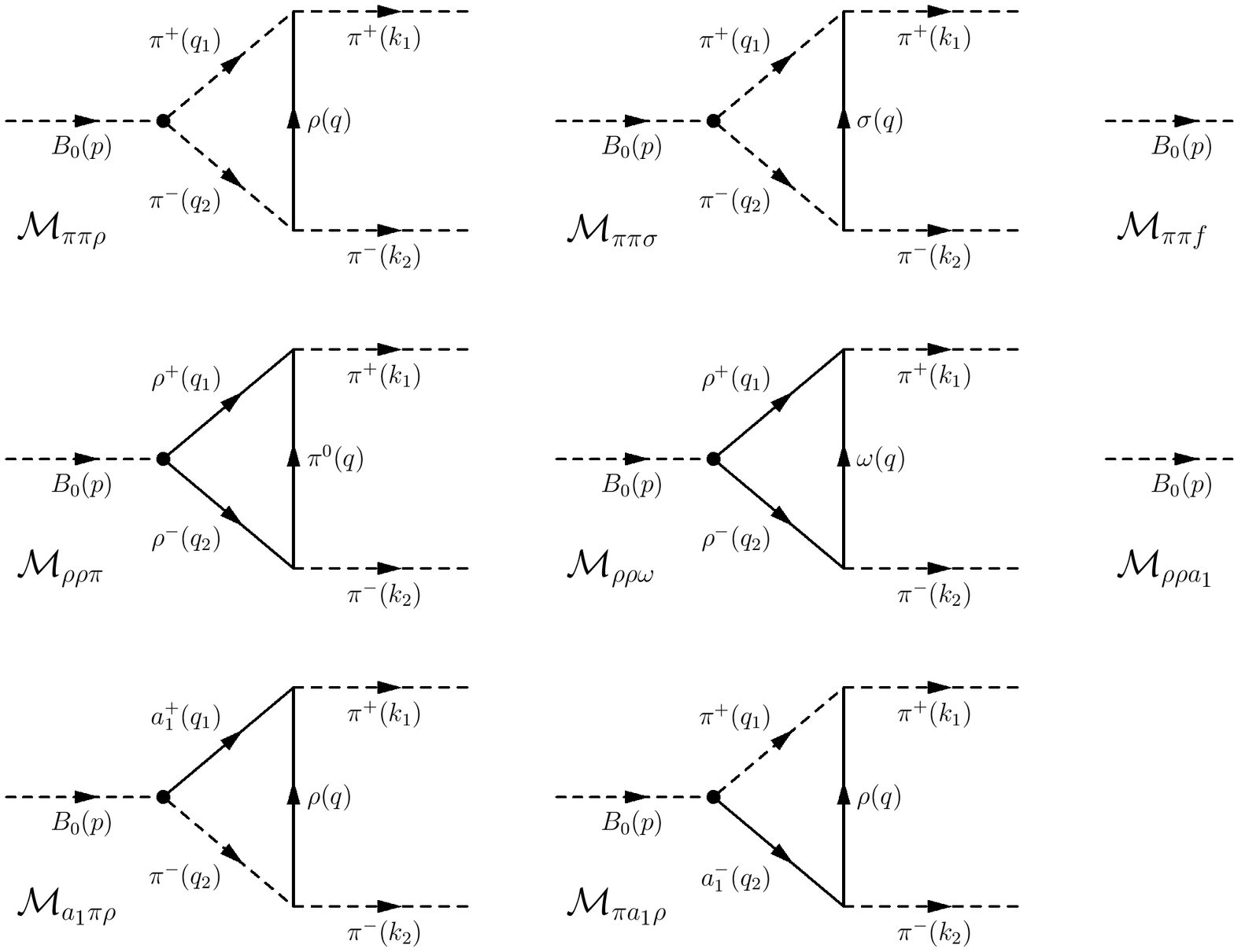}
\caption{Feynman diagrams for $\bar B_0 \to \pi^+ \pi^-$ decay
coming from rescattering of $\pi \pi$ via exchanges of $\rho$, 
$\sigma, f_0$ , $\rho \rho$ rescattering via exchanges of $\pi$, 
$\omega$, $a_1$ and 
$a_1 \pi$ rescattering via exchange of $\rho$. }
\end{figure}
The absorptive parts of amplitudes are obtained when the cut is done over the 
 intermediate states $\pi \pi$, $\rho \rho$ and $a_1 \pi$ 
as schematically given in  Fig. 2. In our further formulas we denote momenta 
of particles as given in 
Fig. 2.
\begin{figure}
\begin{center}
\includegraphics[width=12cm]{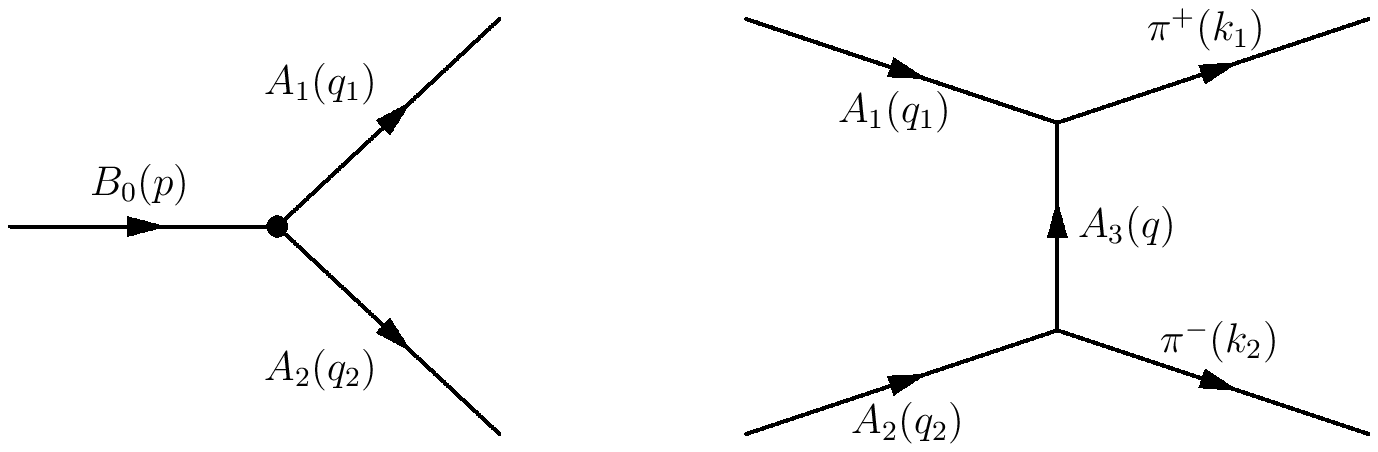}
\caption{The absorptive parts of amplitudes are obtained when the cut is done over the 
 intermediate states $A_1$ and $A_2$.}
\end{center}
\end{figure}
The couplings describing the strong 
interactions of light mesons in these diagrams are all far of mass shell. 
In the approach of 
 \cite{cheng1,ablikim} the additional form factor was included. Its role is 
to take care of the off-mass 
shell effects \cite{gort}:
\begin{equation}
F(y,M_{3})= \left(\frac{\Lambda^2 - M_{3}^2}{\Lambda^2 - t(y)}\right)\,,
\end{equation}
where $t(y) =(q_1 - k_1)^2 $, $\Lambda =
M_3 + \Lambda_{QCD}$ and $M_3$ is the mass of exchanged particle $A_3$ (see Fig. 2). 
We take $\Lambda_{\rm QCD}= 0.3\pm 0.05$ GeV.  
Following the contributions given in Fig. 1, we determine the absorptive parts of the amplitudes

\begin{eqnarray}
&& {\cal M}^{\pi \pi \rho}_A = -{\cal A}_\pi  
g_{\rho \pi \pi}^2\frac{\lambda^{1/2}(m_B^2, m_\pi^2, m_\pi^2)}{32 \pi m_B^2}
\int_{-1}^{1} d y C_{\pi}(y) 
\frac{F^2(y,m_{\rho})}{2m_\pi^2 - 2 S(y)-m_\rho^2}\,,
\label{a1}
\\
&&{\cal M}^{\pi \pi \sigma}_A = -{\cal A}_\pi  
4G_s^2\frac{\lambda^{1/2}(m_B^2, m_\pi^2, m_\pi^2)}{32 \pi m_B^2}
\int_{-1}^{1} dy  
\frac{F^2(y,m_{\sigma})}{2m_\pi^2 - 2 S(y)-m_\sigma^2}\,,
\label{a2}
\\
&& {\cal M}^{\pi \pi f}_A = -{\cal A}_\pi  
4G^{\prime 2}_{s}\frac{\lambda^{1/2}(m_B^2, m_\pi^2, m_\pi^2)}{32 \pi m_B^2}
\int_{-1}^{1} dy  
\frac{F^2(y,m_f)}{2m_\pi^2 - 2 S(y)-m_f^2}\,,
\label{a3}
\\
&& {\cal M}^{\rho\rho\pi}_A = -{\cal A}_\rho g_{\rho \pi \pi}^2
\frac{\lambda^{1/2}(m_B^2, m_\rho^2, m_\rho^2)}{32 \pi m_B^2}\int_{-1}^{1} 
d y \frac{F^2(y,m_{\pi})}{m_\pi^2 + m_\rho^2 - 2 S(y)-m_\pi^2}\times
\nonumber \\
&& \left((m_B + m_\rho)A_1(m_\rho^2) C_{\rho,1}(y)- 2 A_2(m_\rho^2) /(m_B + m_\rho)
C_{\rho,2}(y)\right)\,,
\label{a4}
\\
&& {\cal M}^{\rho\rho\omega}_A = -2{\cal A}_\rho \left(\frac{4C_{VVP}}{f_\pi}\right)^2
\frac{\lambda^{1/2}(m_B^2, m_\rho^2, m_\rho^2)}{32 \pi m_B^2}\int_{-1}^{1} 
d y \frac{F^2(y,m_{\omega})}{m_\pi^2 + m_\rho^2 - 2 S(y)-m_\omega^2}\times
\nonumber \\
&& \left((m_B + m_\rho)A_1(m_\rho^2) C_{\rho,3}(y)- 2 A_2(m_\rho^2) /(m_B + m_\rho)
C_{\rho,4}(y)\right)\,,
\label{a5}
\\
&& {\cal M}^{\rho\rho a_1}_A = -2 {\cal A}_\rho G_{AVP}^2
\frac{\lambda^{1/2}(m_B^2, m_\rho^2, m_\rho^2)}{32 \pi m_B^2}\int_{-1}^{1} 
d y \frac{F^2(y,m_{a_1})}{m_\pi^2 + m_\rho^2 - 2 S(y)-m_{a_1}^2}\times
\nonumber \\
&& \left((m_B + m_\rho)A_1(m_\rho^2) C_{\rho,5}(y)- 2 A_2(m_\rho^2) /(m_B + m_\rho)
C_{\rho,6}(y)\right)\,,
\label{a6}
\\
&& {\cal M}^{\pi a_1\rho}_A=- {\cal A}_{a_1,1} {\sqrt 2} G_{AVP} g_{\rho \pi \pi}  
\frac{\lambda^{1/2}(m_B^2, m_\pi^2, m_{a_1}^2)}{32 \pi m_B^2}\times
\nonumber \\
&&\int_{-1}^{1} d y  C_{a_1,1}(y)
\frac{F^2(y,m_\rho)}{2m_\pi^2- 2 S(y)-m_\rho^2}\,,
\label{a7}
\\
&& {\cal M}^{a_1\pi\rho}_A = - {\cal A}_{a_1,2} {\sqrt 2}G_{AVP} g_{\rho \pi \pi} 
\frac{\lambda^{1/2}(M_B^2, m_\pi^2, m_{a_1}^2)}{32 \pi m_B^2}\times
\nonumber\\
&&\int_{-1}^{1} d y  C_{a_1,2} (y)
\frac{F(y,m_\rho)^2}{m_\pi^2 + m_{a_1}^2 - 2 S(y)-m_\rho^2}\,,
\label{a8}
\end{eqnarray}
where with $C$ stands for the functions of momenta defined in Appendix A, while 
$S(y)$ is the scalar product:
\begin{equation}
S(y) = k_1\cdot q_1 = k_{10} E_1 -|\vec k_1 ||\vec q_1| y\, 
\label{S}
\end{equation}
and $y = cos (\vec k_1,\vec q_1) $. 
We use  
$|\vec q_1|^2= \frac{1}{4 m_B^2} \lambda (m_B^2, M_1^2, M_2^2)$,  
$|\vec k_1|^2= \frac{1}{4 m_B^2} \lambda ( m_B^2, m_\pi^2, m_\pi^2)$
and $E_1^2 = |\vec q_1|^2 + M_1^2$ and $k_{10}^2 = |\vec k_1|^2 + m_\pi^2$. 
Here $M_i$ stands for the masses of intermediate particles  $A_i$  and 
 $\lambda(a,b,c) = a^2 +b^2 + c^2 -2 ab - 2 cb - 2 ac$ as usual. 

\section{DISCUSSION}

 After numerical evaluation\footnote{Numerical results were obtained with the
help of the  computer program FeynCalc \cite{FeynCalc}. }
of these integrals we present our results in 
Table 1.  We give values of the absorptive parts of the amplitude for three different values of the 
scale $\Lambda =0.25; 0.3; 0.35$ GeV. As seen from the table these amplitudes 
are sensitive to  the choice of 
this parameter.
\begin{table}
\begin{center}
\begin{tabular}{|c||c|c|c|}
\hline
$ \enspace$ & $\Lambda= 0.25\enspace {\rm GeV}$ & $\Lambda= 0.30\enspace 
 {\rm GeV}$ & \enspace $\Lambda= 0.35 \enspace{\rm GeV}$ \\
\hline\enspace
$\pi \pi (\rho)  $    & $13.3+0.5i$      & $17.3+0.6i$    & $21.8+0.8i $  \\
$\pi \pi (\sigma)$    & $-0.5-0.02i$    & $-0.6-0.02i$   & $-0.8-0.03i$  \\
$\pi \pi (f_0)$       & $-0.03-0.001i$  & $-0.05-0.002i$ & $-0.06-0.002i$\\
\hline	
$ \Sigma_{\pi \pi}$   & $12.5+0.5i$     & $16.7+0.6i$    & $21+0.8i$    \\
\hline
$\rho \rho (\pi)$     & $-1.7-0.06i$ 	& $-2.2-0.08i$  & $-2.8-0.1i$   \\
$\rho \rho (\omega) $ & $5.5+0.2i$ 	& $7.7+0.3i $  & $10.3+0.4i$   \\
$\rho \rho (a_1)$     & $-0.9-0.03i$    & $-1.4-0.05i$ & $-1.6-0.06i$	 \\
\hline
$ \Sigma_{\rho \rho}$ & $2.8+0.1i$ 	& $4.3+0.2i$   & $5.9+0.2i$	\\
\hline
$a_1^- \pi^+ (\rho^0)$& $5.6+0.2i$	& $7.5+0.3i$   & $9.5+0.3i$	\\
$a_1^+ \pi^- (\rho^0)$& $1.9+0.1i$  	& $2.5+0.1i$   & $3.2+0.1i$	\\ 
\hline
\end{tabular}
\caption{ The absorptive parts of amplitudes coming from the diagrams 
${\cal M}_A^{i}\times 10^{-7} V_{ub} [\rm{GeV}]$ 
given in Fig. 1.}
\label{Table-check}
\end{center}
\end{table}
It is important to note that the relative sign of these contributions 
cannot be completely determined  \cite{ablikim}.  
By assuming that strong couplings do not have any phases,  
the sum of contributions 
 coming from the $\pi\pi \to \pi \pi$ rescattering  is 
then 
$ \Sigma_{\pi \pi}=(1.7+0.06i)\times 10^{-6} V_{ub}$ GeV, which for $|V_{ub}|= 0.00439$
gives $ |\Sigma_{\pi \pi}|=7.5 \times 10^{-9}$ GeV (for $\Lambda = 0.3$ GeV)  . 
It is interesting that the  exchanges of scalar mesons give very small 
contributions. 
The contribution of $\rho \rho$ 
intermediate states with the exchanges of $\pi^0$, $\omega$ and $a_1$ 
is about four times smaller than the total  $\pi^+\pi^-$ intermediate state 
contribution. 
Among these the effect of the  $\omega$ exchange is important. This contribution  
was not considered in \cite{cheng1}. 
The contributions of $a_1 \pi$ intermediate states might be  significant, 
close in size to the leading $ \pi^+ \pi^-$ elastic-rescattering effect. 
Then in the best case (by summing the contributions given in  Table 1, 
all with the positive signs)  we 
can give an upper  value for the absorptive part
of the amplitude ($\Lambda = 0.3$ GeV):
\begin{equation}
|{\cal M}_{A} (\bar B^0 \to \pi^+ \pi^-)| \leq 1.7 \times 10^{-8} {\rm GeV}. 
\label{c1a}
\end{equation}

This  value  is very close in size to the 
short distance amplitude discussed in \cite{cheng1} (Eqs. (5.14)). 
On the other hand, for the certain choice of the strong couplings phases,  
the calculated contributions might almost cancel each other, leading  
to the disappearance of the absorptive part of FSI amplitude.

 In the case of $\bar B^0 \to \pi^0 \pi^0$
the absorptive part of amplitude comes from the same FSI and the upper bound is  
$| {\cal M}_A(\bar B^0 \to \pi^0 \pi^0)| 
\leq  1.4 \times 10^{-8} {\rm GeV}$,  ($\Lambda = 0.3$ GeV). 
Note that there are no contributions coming from the exchanges of 
neutral mesons as $\sigma$, $f_0$ in the case of $\pi^+ \pi^- \to \pi^0 \pi^0$
 and $\omega$ in $ \rho^+ \rho^- \to \pi^0 \pi^0$ mode. 
Comparing this result with short distance amplitude given in \cite{cheng1} 
(Eqs. (5.14))
we  see that the effect we discuss 
might enhance  the amplitude by a factor of 2. However, the corresponding 
branching ratio 
 is still too small in comparison with the 
experimental result.

In order to estimate the effects of this leading FSI contribution in 
$B^- \to \pi^- \pi^0$ decay amplitudes one can rely on the isospin 
relation\footnote{Note that we have used the Feynman diagram convention for the $\pi^0 \pi^0$ amplitude 
as in \cite{kamal}.}

\begin{equation}
A(\bar B^0 \to \pi^+ \pi^-) - A(\bar B^0 \to \pi^0 \pi^0) = - {\sqrt 2} 
 A(B^- \to \pi^0 \pi^-).
\label{ai}
\end{equation}

We find that the absorptive part from $\pi \pi$ (elastic rescattering)
and quasi-elastic FSI $\rho \rho$ via the $t$-channel $\pi$, $a_{1}$,
$\omega$-exchange  contributions might be important for 
$B\to\pi \pi$  amplitudes.  
Here we point out that the absorptive part of the 
$\bar B^0 \to \pi^+ \pi^-$  amplitude produces the phase 
of the tree amplitude of \cite{Wolfenstein,wu} 
while the absorptive part of the $\bar B^0 \to \pi^0 \pi^0$ amplitude determines the 
color-suppressed phase of the amplitude in \cite{Wolfenstein,wu}. In a recent paper \cite{Wolfenstein} 
it was shown that it is possible 
to determine the strong phase
separately for the tree, color-suppressed, and penguin amplitudes from
the current BaBar and Belle measurements on $B \to \pi\pi$
branching ratios and CP asymmetries. The results show that the relative
phase between the tree and color-suppressed amplitudes $\delta_{T}-\delta_{C}$
is rather small. Since we found the strong phase coming from calculated FSI effect 
for  $\pi^+ \pi^-$ (tree amplitude) and 
$\pi^0 \pi^0$ (color-suppressed amplitude) to be almost of the same size, 
we can confirm the results of the phenomenological study given in Ref.  
\cite{Wolfenstein}.

Our calculations contain only information on the absorptive part of 
amplitudes indicating sources of uncertainties.  
One can in principle determine the dispersive parts 
of amplitudes, but due to many uncertainties we do not pursue in calculating  
 these effects. As noticed in \cite{charm-pen,charm-pen1} these contributions 
are expected to be of  similar size as the absorptive
parts of amplitudes for both $\pi^+ \pi^-$ and $\pi^0 \pi^0$ decay modes.

Recently the authors of Ref. \cite{deandrea} estimated
the effects of final state interactions using the Regge model. 
This analysis shows that the long distance charming penguins do not play 
important role. However, the long distance 
effects due to the light meson rescattering
are very important in obtaining correct rates for 
$B \to \pi \pi$ decays \cite{deandrea}, 
in agreement with the result of our calculation. 

In Ref. \cite{wu}, using the $SU(3)$ symmetry relations, it was found that 
in $ B \to \pi \pi$  decays  
the ratio of the color-suppressed and tree amplitudes  is very large. 
Our calculations, obtained within a very different framework, confirm this 
finding.

\section{SUMMARY}

We can briefly summarize our results:\\

1) The absorptive parts of amplitudes in $B \to \pi \pi$ decays are calculated 
using the rescattering of 
 $\pi \pi$ via exchange of  $\rho$, 
$\sigma$, $f_0$; 
$\rho \rho$ rescattering via exchange of $\pi$, $\omega$, $a_1$ and 
contributions of the  $a_1 \pi$ rescattering 
via exchange of $\rho$.\\

2) Although our results suffer from many uncertainties  due to unknown 
relative  phases and the dependence on 
the parameter $\Lambda$, we can say that our study shows the 
importance of the charmless 
final state interactions in $B \to \pi \pi$ decays.
Both  the $\bar B^0 \to \pi^+ \pi^-$ and $\bar B^0 \to \pi^0 \pi^0$ amplitudes 
might get significant contributions from absorptive parts of the FSI amplitudes. \\ 

3) Our result shows that the relative
phase between  the tree and color-suppressed amplitude $\delta_{T}-\delta_{C}$
is rather small and in agreement with the results of previous 
phenomenological studies. 

\vspace{1cm}

\centerline{\bf Acknowledgments}

S.F. thanks Alexander von Humboldt foundation for financial support and A. J. Buras for his warm hospitality 
during her stay at the Physik Department, TU M\" unchen, where part of this work has been done. 
The work of S.F. and A.P. has been supported in part by the Ministry of Higher Education,
Science and Technology of the Republic of Slovenia.

\section{ APPENDIX A}

The  functions of momenta  $C$ are (momenta $q_i$, $k_i$ and $q$ are defined in Fig. 2): 

\begin{equation}
C_{\pi} (y) =  (q_1 + k_1)^\alpha (q_2 + k_2)^\beta(- g_{\alpha \beta} + 
q_\alpha q_\beta/{m_i^2})=2\left(m_\pi^2-m_b^2+S(y)\right)\,,
\label{c1}
\end{equation}

\begin{eqnarray}
&&C_{\rho,1}(y) = (- g_{\alpha \beta} + q_{1\alpha} q_{1\beta}/m_\rho^2)
(- g^{\alpha \delta} + q_2^{\alpha} q_2^{\delta}/m_\rho^2)
(2 k_{1} - q_1)^{\beta} (2 k_2 -q_2)_{\delta} \nonumber\\
&&= \frac{2}{m_\rho^4}\left(-2m_\pi^2 m_\rho^4 + m_B^2 m_\rho^4
+ 2S(y)\left(S(y)-m_B^2\right) m_\rho^2 + m_B^2 S(y)^2\right)\,,
\label{c2}
\end{eqnarray}

\begin{eqnarray}
&&C_{\rho,2}(y) = (k_1 + k_2)_{\alpha}(k_1 + k_2)_{\gamma}
(- g^{\alpha \beta}  + q^{\alpha}_1
q^{\beta}_1/m_\rho^2)\nonumber\\
&&
(2 k_1 - q_1)_{\beta} (-g^{\gamma \delta} + q_2^\gamma q_2^\delta/m_\rho^2)
(2 k_2 - q_2)_{\delta}
=\frac{m_B^4}{m_\rho^4}\left(m_\rho^2 - S(y)\right)^2\,,
\label{c3}
\end{eqnarray}

\begin{eqnarray}
&& C_{\rho,3}(y) = (- g_{\alpha \sigma} + q_{2\alpha}
q_{2\sigma}/m_\rho^2)
(- g_{\alpha \sigma'} + q_{1\alpha} q_{1\sigma'}/m_\rho^2) \nonumber\\
&&(- g_{\gamma \gamma'} + (k_1- q_1)_\gamma  (k_1-
q_1)_{\gamma'}/m_\omega^2)
\epsilon^{\kappa \sigma \rho \gamma}
\epsilon^{\kappa' \sigma' \rho' \gamma'}  
 q_{2 \kappa} q_{1 \kappa'} (k_1 -q_1)_{\rho} (k_1 -
 q_1)_{\rho'}\nonumber\\
&& = m_\pi^2\left(m_B^2 - 2 m_\rho^2\right) + m_\rho^2 m_B^2 + 2S(y)\left(S(y)-m_B^2\right)\,,
\label{c4}
\end{eqnarray}

\begin{eqnarray}
&& C_{\rho,4}(y)= (k_1 + k_2)^{\alpha'}(- g_{\alpha' \sigma} + q_{2\alpha'} 
q_{2\sigma}/m_\rho^2)(k_1 + k_2)^\alpha 
(- g_{\alpha \sigma'} + q_{1\alpha} q_{1\sigma'}/m_\rho^2)\nonumber\\ 
&&(- g_{\gamma \gamma'} + (k_1- q_1)_\gamma  (k_1-
q_1)_{\gamma'}/m_\omega^2)
\epsilon^{\kappa \sigma \rho \gamma}
\epsilon^{\kappa' \sigma' \rho' \gamma'} 
 q_{2 \kappa} q_{1 \kappa'} (k_1 -q_1)_{\rho} (k_1 -
 q_1)_{\rho'}\nonumber\\
&& =\frac{m_B^2}{4}\left(\left(m_B^2-4 m_\rho^2 \right)m_\pi^2 
+ m_\rho^2 m_B^2- 2 S(y)\left(m_B^2 -2 S(y)\right)  \right)\,,
\label{c5}
\end{eqnarray}

\begin{eqnarray}
&& C_{\rho,5(y)}= ( - g_{\alpha \gamma} + q_{2\alpha} q_{2 \gamma}/m_\rho^2)
(- g^{\gamma \rho} + q^{\rho} q^{\gamma}/m_{a1}^2)(- g^{\alpha} _{\rho} + q_{1\rho} q_1^{\alpha}/m_\rho^2)\nonumber\\
&& = -\frac{1}{4 m_\rho^4 m_{a1}^2} \left(
2 m_\rho^4 \left(6 m_{a1}^2 +m_B^2\right) - 4 m_\pi^2 m_\rho^4\right.\nonumber\\
&& \left.-4 m_\rho^2 \left(m_{a1}^2 m_B^2+ S(y)\left(m_B^2 -S(y)\right)\right)
+m_{a1}^2 m_B^4 + 2 m_B^2 S(y)^2\right)\,, 
\label{c6}
\end{eqnarray}

\begin{eqnarray}
&& C_{\rho,6(y)}= (k_1 + k_2)_{2\beta} (k_1 + k_2)^{\alpha}( - g_{\alpha \gamma} + q_{2\alpha} q_{2 \gamma}/m_\rho^2)
(-g^{\gamma \rho} + q^{\rho} q^{\gamma}/m_{a1}^2)
\nonumber\\&&
(- g^{\beta} _{\rho} + q_{1\rho} q_1^{\beta}/m_\rho^2)
= - \frac{m_B^2 }{8  m_\rho^4m_{a1}^2} \left(2 m_\rho^4\left(4 m_{a1}^2 + m_B^2\right)
\right.\nonumber\\&&\left.
- 2m_B^2 m_\rho^2 \left(3 m_{a1}^2 + 2 S(y) \right)+ m_{a1}^2 m_B^4 + 2 m_B^2 S(y)^2\right). 
\label{c7}
\end{eqnarray}

\begin{eqnarray}
&&C_{a_1,1} (y) = (2 q_1 +q_2)^\alpha ( - g_{\alpha \beta} + q_{2\alpha} q_{2
\beta}/m_{a_1}^2)
( - g^{\beta \delta} + q_{\beta} q_{\delta}/m_{\rho}^2 )(q_1 + k_1)_{\delta}\nonumber\\
&&=\frac{-1}{2m_{a1}^2}\left(m_\pi^4+m_\pi^2\left(2S(y)-3m_B^2-2m_{a1}^2\right)+m_{a1}^4
\right. \nonumber\\&&\left. +2m_B^2\left(m_B^2-S(y)\right)-m_{a1}^2
\left(3m_B^2+2S(y)\right)\right)\,,
\label{c8}
\end{eqnarray}

\begin{eqnarray}
&&C_{a_1,2} (y) = (2 q_1 +q_2)^\alpha ( - g_{\alpha \beta} + q_{1\alpha} q_{1
\beta}/m_{a_1}^2)
( - g^{\beta \delta} + q_{\beta} q_{\delta}/m_{\rho}^2)(q_2 + k_2)_{\delta}\nonumber\\
&&=\frac{-1}{2m_{a1}^2}\left(m_\pi^4+m_\pi^2\left(S(y)-2m_B^2-2m_{a1}^2\right)+m_{a1}^4
\right. \nonumber\\&&\left.
+m_B^4-m_B^2S(y)-m_{a1}^2\left(m_B^2+S(y)\right)\right)\,.
\label{c9}
\end{eqnarray}

\end{document}